\documentclass[twocolumn,english,preprintnumbers,amsmath,amssymb]{revtex4}
\usepackage[latin9]{inputenc}
\usepackage{amsmath}
\usepackage{graphicx}
\usepackage{amssymb}

\makeatletter
\@ifundefined{textcolor}{}
{%
 \definecolor{BLACK}{gray}{0}
 \definecolor{WHITE}{gray}{1}
 \definecolor{RED}{rgb}{1,0,0}
 \definecolor{GREEN}{rgb}{0,1,0}
 \definecolor{BLUE}{rgb}{0,0,1}
 \definecolor{CYAN}{cmyk}{1,0,0,0}
 \definecolor{MAGENTA}{cmyk}{0,1,0,0}
 \definecolor{YELLOW}{cmyk}{0,0,1,0}
 }

\makeatletter
\@ifundefined{textcolor}{}
{%
 \definecolor{BLACK}{gray}{0}
 \definecolor{WHITE}{gray}{1}
 \definecolor{RED}{rgb}{1,0,0}
 \definecolor{GREEN}{rgb}{0,1,0}
 \definecolor{BLUE}{rgb}{0,0,1}
 \definecolor{CYAN}{cmyk}{1,0,0,0}
 \definecolor{MAGENTA}{cmyk}{0,1,0,0}
 \definecolor{YELLOW}{cmyk}{0,0,1,0}
 }

\usepackage{dcolumn}
\usepackage{bm}

\newcommand{\comment}[1]{}
\newcommand{\li}{$^{6}$Li }
\newcommand{\yb}{$^{174}$Yb }

\@ifundefined{definecolor}
 {\@ifundefined{definecolor}
 {\@ifundefined{definecolor}
 {\@ifundefined{definecolor}
 {\usepackage{color}}{}
}{}
}{}
}{}

\makeatother

\usepackage{babel}

\makeatother

\usepackage{babel}

\makeatother

\usepackage{babel}

\makeatother

\usepackage{babel}

\begin{document}

\title{Dynamics of Feshbach Molecules in an Ultracold Three-Component Mixture}

\author{Alexander Y. Khramov}

\author{Anders H. Hansen}

\author{Alan O. Jamison}

\author{William H. Dowd}

\author{Subhadeep Gupta}

\affiliation{Department of Physics, University of Washington, Seattle WA 98195}

\date{\today}
\begin{abstract}
We present investigations of the formation rate and collisional stability
of lithium Feshbach molecules in an ultracold three-component mixture
composed of two resonantly interacting fermionic $^{6}$Li spin states
and bosonic $^{174}$Yb. We observe long molecule lifetimes ($>\,100\,$ms)
even in the presence of a large ytterbium bath and extract reaction
rate coefficients of the system. We find good collisional stability
of the mixture in the unitary regime, opening new possibilities for
studies and probes of strongly interacting quantum gases in contact
with a bath species.
\end{abstract}

\maketitle
Magnetic Feshbach resonances allow precise control of collisional
properties, making them a key tool in ultracold atom systems. They
have been extensively used to study ultracold molecules, as well as few- and
many-body physics \cite{chin10}. Two-component Fermi gases near a
Feshbach resonance provide excellent opportunities to study strongly
interacting quantum systems \cite{gior08}. This is possible due to
the remarkable collisional stability of the atom-molecule mixture
on the positive scattering length side of the resonance \cite{joch03,cubi03},
attributed largely to Fermi statistics \cite{esry01,petr04}. Extending
the system to three-component mixtures in which only two are resonantly
interacting \cite{spie09} offers the exciting possibility of modifying
or probing pairing dynamics by selective control of the third component.
A third-component may also be used as a coolant bath for
exothermic molecule-formation processes, provided that inelastic processes with
the bath are negligible. In the context of many-body physics, a third
non-resonant component can be useful as a microscopic probe
of superfluid properties \cite{spie09,targ10}, as a stable bath for studies of
non-equilibrium phenomena \cite{robe09}, or as an accurate thermometer
of deeply degenerate fermions \cite{nasc10}.

Collisional stability of Feshbach molecules in the absence of Fermi
statistics becomes a crucial question for multi-component
mixtures \cite{dinc08,zirb08,spie09}. A recent theoretical analysis of such
mixtures suggests a possibility for enhanced molecule formation rates
with good collisional stability \cite{dinc08}. Enhanced atom loss
has been observed near a \li \emph{p}-wave resonance in the presence
of a $^{87}$Rb bath \cite{deh08}, while a small sample of the probe
species $^{40}$K has been found to be stable within a larger strongly
interacting \li sample \cite{spie09}.

In this paper we investigate a mixture composed of two resonantly
interacting spin states of fermionic \li immersed in a large sample
of bosonic $^{174}$Yb atoms. While the Li interstate interactions
are arbitrarily tunable by means of an $s$-wave Feshbach resonance
at $834\,$G \cite{diec02}, the interspecies interactions between
Li and Yb are constant and small \cite{ivan11}. We present the first observations
of formation and evolution of Feshbach molecules in a bath of a
second atomic species. In the unitary regime, we observe good collisional
stability of the mixture with elastic interactions dominating over inelastic
losses. We extract the reaction rate constants from a classical
rate equations model of the system.

Our experimental procedure has been described in earlier work \cite{hans11}.
Briefly $3\!\times\!10^{6}$ atoms of \yb in the $^{1}S_{0}$ state
and up to $4\times10^{4}$ atoms of \li, distributed equally between
the two $^{2}S_{1/2}$, $F\!=\!\frac{1}{2}$ states (denoted Li$|1\rangle$,
Li$|2\rangle$), are loaded from magneto-optical traps into a crossed-beam
optical dipole trap. We then perform forced evaporative cooling on Yb
to the final trap depth $U_{{\rm Yb(Li)}}=15(55)\,\mu$K, with  mean trap frequency
$\bar{\omega}_{{\rm Yb(Li)}}=2\pi\times0.30(2.4)\,$kHz \cite{odt}, during which
Li is cooled sympathetically by Yb. Following evaporation, the mixture is held at constant trap depth
to allow inter-species thermalization. With a time constant of $1\,$s
the system acquires a common temperature $T_{{\rm Yb}}\,=\, T_{{\rm Li}}\,=\,2\,\mu$K
with atom number $N_{{\rm Yb(Li)}}=2\times10^{5}(3\times10^{4})$.
This corresponds to $T_{{\rm Li}}/T_{F}\simeq0.4$, and $T_{{\rm Yb}}/T_{C}\simeq2.5$,
where $T_{F}$ is the Li Fermi temperature and $T_{C}$ is the Yb
Bose-Einstein condensation temperature \cite{bectf}.

\begin{figure}[b]
\includegraphics[width=0.5\textwidth]{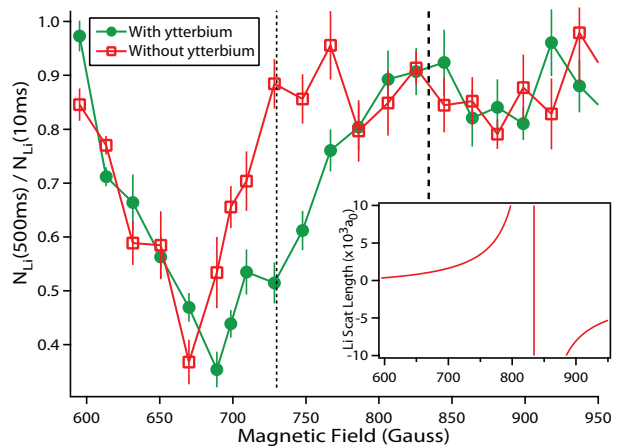}\vspace{-0.5mm}
\caption{\label{lspec} Li atom loss spectroscopy in the presence (filled circles)
and absence (open squares) of an Yb bath near the \li $834\,$G Feshbach resonance (inset).
We plot the number of Li atoms after $500\,$ms of evolution normalized to that at $10\,$ms.
The thick dashed line indicates the resonance center and the
thin dashed line indicates the magnetic field at which $\epsilon_{B}=k_{B}T_{\rm Li}$ for
the initial conditions.}
\label{fig:lspec}
\end{figure}

After this initial preparation, we ramp up the magnetic field to a desired value
and observe the system after a variable hold time. For fields in the vicinity of the Feshbach resonance,
there is a field-dependent number loss and heating for the Li cloud during
the $20\,$ms ramp time, resulting in $T_{{\rm Li}}$ rising to as
high as $4.5\,\mu$K. At this point, the density-weighted average density $\langle n_{{\rm Yb(Li)}}\rangle$
is $2.6(0.35)\times10^{13}\,$cm$^{-3}$. For interrogation in the absence
of the bath, Yb is removed from the trap with a $1\,$ms light pulse
resonant with the $^{1}S_{0}\rightarrow{^{1}P_{1}}$ transition \cite{ybblast}.
Atom number and temperature are monitored using absorption
imaging for both species after switching off the magnetic field.

We first present our results on atom loss spectroscopy near the Feshbach
resonance (see Fig.\,\ref{fig:lspec}). The atom-loss maximum obtained in
the absence of Yb has been observed previously \cite{diec02} and can be explained
as a result of the formation and subsequent decay of shallow lithium
Feshbach dimers \cite{joch03,cubi03,chin04,kokk04,zhan11} which form only on the positive $a$ side of the resonance. Here $a$ denotes the Li$|1\rangle$-Li$|2\rangle$ scattering length. In the presence of the
Yb bath, the loss feature is shifted and broadened. We interpret the behavior of the mixture in
terms of five chemical processes:
\begin{align*}
\mathrm{Li|1\rangle+Li|2\rangle+Li}\hspace{4mm} & \rightleftharpoons & \mathrm{Li_{2}^{\mathnormal{s}}+Li\hspace{6.2mm}(+\epsilon_{B})}\tag{I}\\
\mathrm{Li_{2}^{\mathnormal{s}}+Li}\hspace{4mm} & \rightarrow & \mathrm{Li_{2}^{\mathnormal{d}}+Li\hspace{6.2mm}(+\epsilon_{D})}\tag{II}\\
\mathrm{Li|1\rangle+Li|2\rangle+Yb}\hspace{4mm} & \rightleftharpoons & \mathrm{Li_{2}^{\mathnormal{s}}+Yb\hspace{5mm}(+\epsilon_{B})}\tag{III}\\
\mathrm{Li_{2}^{\mathnormal{s}}+Yb}\hspace{4mm} & \rightarrow & \mathrm{Li_{2}^{\mathnormal{d}}+Yb\hspace{5mm}(+\epsilon_{D})}\tag{IV}\\
\mathrm{Li|1\rangle+Li|2\rangle+Yb}\hspace{4mm} & \rightarrow & \mathrm{Li_{2}^{\mathnormal{d}}+Yb\hspace{5mm}(+\epsilon_{D})}\tag{V}
\end{align*}
Forward process (I) corresponds to a three-body collision event which
produces a shallow Feshbach dimer (denoted $\mathrm{Li_{2}^{\mathnormal{s}}}$)
accompanied by the release of the dimer binding energy $\epsilon_{B}=\frac{\hbar^{2}}{2m_{{\rm Li}}a^{2}}$.
Li denotes a \li atom in either of the two spin states. Process (II)
corresponds to two-body loss to a deeply bound dimer (denoted $\mathrm{Li_{2}^{\mathnormal{d}}}$)
with binding energy $\epsilon_{D}$. Processes (III) and (IV) are
similar to (I) and (II) with the spectator atom being Yb rather than
Li \cite{prociv}. Process (V) corresponds to direct three-body loss to a deeply-bound
molecule. Processes (II, IV, V) always result in particle loss from
the trap since $\epsilon_{D}\gg U_{\rm Li}$. Vibrational relaxation due
to collisions between ${\rm Li}_{2}^{\mathnormal{s}}$
Feshbach molecules may contribute at the lowest fields, but has a
negligible rate for the fields at which we perform our analysis
\cite{joch03,zhan11}. We have experimentally checked that direct
three-body loss processes to deeply-bound states involving three Li atoms
as well as those involving one Li atom and two Yb atoms are negligible for
this work \cite{threebody}. Three-body losses involving Yb atoms alone have a small effect
\cite{taka03} and are taken into account in our analysis.

\begin{figure}
\includegraphics[width=0.5\textwidth]{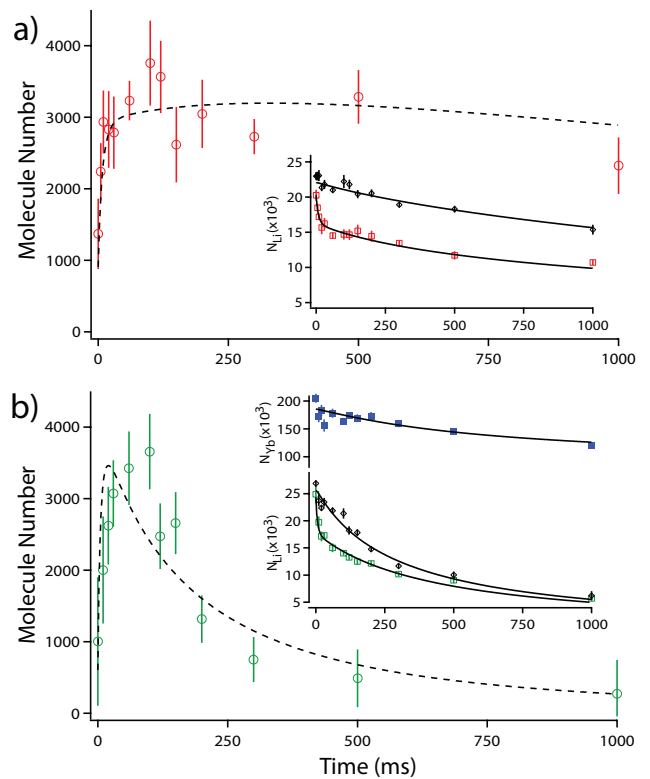}
\vspace{-0.5cm}
\caption{\label{molevol} Evolution of Li Feshbach molecule number at 709 G
without (a) and with (b) an Yb bath. The numbers are obtained by comparing
Li atom numbers (insets) ramped across resonance (diamonds)
or not (open squares) as described in the text. Lower inset also shows
Yb number (filled squares). The curves are fits with a rate equations-based
model.}
\label{fig:molevol}
\end{figure}

In the absence of Yb, only processes (I) and (II) contribute. If we
neglect loss process (II), the atom-molecule mixture approaches an
equilibrium, characterized by an equality of the forward and reverse
rates and an equilibrium molecule fraction
$\frac{2N_{m}}{N_{\rm Li}+2N_{m}}=\left(1+\frac{e^{-\epsilon_{B}/k_{B}T}}{\phi_{\mathrm{Li}}}\right)^{-1},$
where $N_m$ is the molecule number and $\phi_{{\rm Li}}$ is the phase space density for each spin
component in the ground state of the trap \cite{chin04,kokk04,zhan11}. The timescale
for achieving equilibrium depends on the three-body rate constant
$L_{3}$ for process (I), which scales with the scattering length
as $a^{6}$, whereas rate constant $L_{2}$ for process (II) scales
as $a^{-3.3}$ \cite{petr03,dinc05}. The shape of the loss spectrum
can thus be qualitatively explained by noting that the dimer formation
rate increases with magnetic field while equilibrium dimer fraction
and molecule decay rate decrease. The large rate for process (I) at high
fields close to resonance ensures equilibrium molecule fraction at all times.
Broadly speaking, the rate-limiting step determining the system evolution is the
molecule formation rate at low fields and decay rate at high fields. The trap
depth also affects the loss spectrum shape, since it determines the magnetic
field range over which the formed shallow dimers remain trapped.

In the presence of Yb, the additional dimer formation (III), dimer
decay (IV), and 3-body loss (V) processes contribute. The observed
loss spectrum is broadened on the higher field side, suggesting that
for our parameters, processes (IV) and/or (V) play an important role
while process (III) does not. The rate constants $L'_{3}$, $L'_{2}$, and $L_{3}^{d}$,
for processes (III), (IV) and (V), have theoretical scalings $a^{4}$,
$a^{-1}$, and $a^{2}$, respectively \cite{dinc08,petr03}. Overall, we see two regimes of
behavior - a lossy one where molecule formation is energetically favored
($\epsilon_{B}>k_{B}T_{\rm Li}$) and a stable one closer to resonance
($\epsilon_{B}<k_{B}T_{\rm Li}$). The criterion $\epsilon_{B}=k_{B}T$ separating
these two regimes is equivalent to $ka=1$ where $\frac{\hbar^2k^2}{2m_{\rm Li}}=k_B T_{\rm Li}$,
i.e., the unitary criterion.

In order to expand upon this qualitative picture, we study the time
evolution of the three-component mixture at representative magnetic
fields in the above two regimes. We are then able to extract quantitative
information for the above processes from a rate-equations
model of the system.

Fig.\,\ref{fig:molevol} shows the Li atom and molecule number evolution
at $709\,$G ($\epsilon_{B}=k_{B}\times8.3\,\mu$K) a field value where modifications
due to the Yb bath are apparent in Fig.\,\ref{fig:lspec}. The number
of Feshbach molecules at a particular field is determined by using
a procedure similar to earlier works \cite{joch03,cubi03}. After variable evolution
time, we ramp the magnetic field with a speed of $40\,$G/ms either
up to $950\,$G, which dissociates the molecules back into atoms that
remain in the trap, or to $506\,$G, which does not. We then rapidly
switch off the magnetic field and image the atomic cloud. The molecule
number is obtained from the number difference in the two images
(see insets in Fig.\,\ref{fig:molevol}).

We see that the presence of Yb alters the molecule decay rate while the formation
rate is unchanged. The Feshbach molecules appear to coexist for a long time ($>100\,$ms)
with the Yb bath, even in the absence of Pauli blocking \cite{dinc08}.
We adapt the recent rate-equations analysis of Feshbach losses in
a Fermi-Fermi mixture \cite{zhan11} to incorporate a third component,
temperature evolution, and trap inhomogeneity. $T_{{\rm Li}}/T_{F}>0.5$
is satisfied throughout the measurement range, allowing a classical
treatment of the Li cloud. We model the density evolutions due to processes (I-V) using:
\begin{eqnarray}
\dot{n}_{m} & = & R_{m}+R'_{m}-L_{2}n_{m}n_{{\rm Li}}-L'_{2}n_{m}n_{{\rm Yb}}\label{eq:nmdot}\\
\dot{n}_{{\rm Li}} & = & -2R_{m}-2R'_{m}-L_{2}n_{m}n_{{\rm Li}}-2L_{3}^{d}n_{{\rm Li}}^{2}n_{{\rm Yb}}\label{eq:nldot}\\
\dot{n}_{{\rm Yb}} & = & -L'_{2}n_{m}n_{{\rm Yb}}-L_{3}^{d}n_{{\rm Li}}^{2}n_{{\rm Yb}}.\label{eq:nydot}
\end{eqnarray}
Here $n_{m}$, $n_{{\rm Li}}$ and $n_{{\rm Yb}}$ are the
densities of shallow dimers $\mathrm{Li_{2}^{\mathnormal{s}}}$, Li atoms and Yb atoms, respectively.
$R_{m}(R'_{m})=\frac{3}{4}L_{3}(L'_{3})n_{{\rm Li}}^{2}n_{{\rm Li(Yb)}}-qL_{3}(L'_{3})n_{m}n_{{\rm Li(Yb)}}$
is the net-rate for molecule production via process (I)((III)). We determine $q$ through the
constraints on the molecule fraction at equilibrium ($R_{m}(R'_{m})=0$). We obtain an upper
bound for $L_{3}^{d}$ by observations at large negative $a$ (described below) which
indicates a negligible effect for the data in Fig.\,\ref{fig:molevol}, allowing us to
set $L_{3}^{d}=0$ for the analysis at $709\,$G.

\begin{figure}
\includegraphics[width=0.5\textwidth]{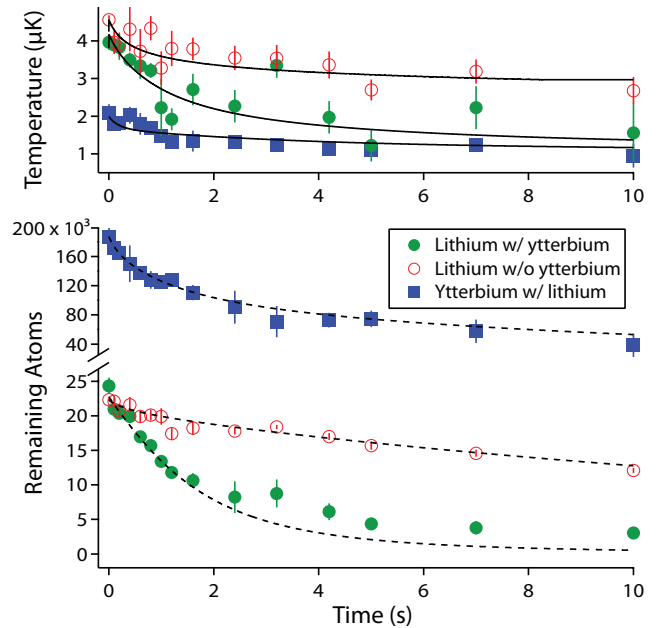}
\vspace{-0.5cm}
\caption{\label{inelast} The evolution of temperature and number at
$810\,$G for Li atomic cloud with Yb (filled circles) and without
(empty circles) and also for Yb in the presence of Li (filled
squares). The curves are fits with a rate equations-based
model.}
\label{fig:inelast}
\end{figure}

The time evolution of $T_{{\rm Li}}$ and $T_{{\rm Yb}}$ are modeled
considering the energy deposition from processes (I) and (III) as
well as heating from the density-dependent loss processes (II), (IV)
and (V) \cite{webe03}. In addition, our model also takes into account
the effects of evaporative cooling \cite{ohar01}, inter-species thermalization
\cite{ivan11}, one-body losses from background gas collisions, and
Yb three-body losses \cite{taka03,webe03}. The Li scattering length at
709$\,$G is $a=1860\, a_{0}$, ensuring rapid thermalization ($<1\,$ms) in the
lithium atom-Feshbach molecule mixture
\cite{petr03}. This allows the assumption of equal temperature $T_{{\rm Li}}$
for lithium atoms and Feshbach molecules. The heating from molecule-formation
at $709\,$G dominates over inter-species thermalization, maintaining
$T_{\rm Li}\simeq4.5\,\mu$K and $T_{\rm Yb}\simeq2\,\mu$K, as observed
in both experiment and model.

The best-fit rate coefficients extracted from the atom data (shown
in the insets) are $L_{3}=(1.4\pm0.3)\times10^{-24}\,{\rm cm}^{6}/$s,
$L_{2}=(1.3\pm0.3)\times10^{-13}\,{\rm cm}^{3}/$s, and $L'_{2}=(2.3\pm0.2)\times10^{-13}\,{\rm cm}^{3}/$s.
$L'_{3}$ is consistent with 0. All reported uncertainties are statistical.
The $L_{3}$ value is consistent with that obtained in \cite{joch03},
after accounting for the slight differences in experimental parameters. Using
$L'_{2}\langle n_{{\rm Yb}}\rangle$ as a measure of the dimer decay
rate, we get $170\,$ms as the lifetime of a Li Feshbach molecule
in the Yb bath.

We now turn to the unitary regime, where we choose 810$\,$G ($ka=+6$,
$\epsilon_{B}=k_{B} \times 0.11\,\mu$K) as our representative field to study the
mixture properties. It is difficult to reliably observe the molecule
number using our earlier method in this regime, so we only monitor
the atoms (see Fig.\,\ref{fig:inelast}). Starting with an inter-species
temperature differential as before, we observe a fast drop in $T_{{\rm Li}}$
in the presence of Yb and clear evidence of inter-species thermalization.
The Li number in the three-component mixture exhibits a long $1/e$ lifetime of $2\,$s,
far larger than at $709\,$G. However this is still an order of magnitude
shorter than that obtained in the absence of Yb.
The interpretation of the decay is not straightforward as both two-body (process (IV)) and
three-body (process (V)) inelastic loss can contribute \cite{du09,spie09}.
The large rate for process (I) in this regime ensures equilibrium molecule fraction at all times.
By fitting to data taken at $935\,$G where $ka=-2$ and process (V) is expected to dominate
inelastic loss, we obtain $L_{3}^{d}=(4.3\pm 0.3)\times10^{-28}\,{\rm cm}^{6}/$s. This sets a lower bound
for $L_{3}^{d}$ at $810\,$G. We fit the first $2.5\,$s of data in Fig.\,\ref{fig:inelast} after fixing $L'_{2}$
to its value scaled from $709\,$G and find $L_{3}^{d}=(9.5\pm 0.5)\times10^{-28}\,{\rm cm}^{6}/$s at $810\,$G.
The slight disagreement in Li atom number at long times may be due to a small ($<10\%$) inequality in our
spin mixture composition, which the model does not take into account.

The qualitative features of both spectra in Fig.\,\ref{fig:lspec} can be theoretically reproduced by using field-dependent
reaction coefficients scaled from our measured values at $709$ and $810\,$G. However, a full quantitative comparison
will need to take into account the theoretical deviations from scaling behavior in the unitary regime
as well as experimental variations in the initial temperature, and is open to future investigation.

By extending the forced evaporative cooling step, lower temperature
mixtures can be produced where bosonic \yb shrinks to a size smaller
than the Fermi diameter of the \li cloud. Such experiments at $834\,$G
yield $T_{{\rm Li}}/T_{F}\simeq0.25$ with $N_{{\rm Yb}}=N_{{\rm Li}}=2.5\times10^{4}$.
Here, the estimated volume of the Yb sample is $\simeq0.3$ of the Li sample volume, compared
to 3.3 in the classical regime. The mixture is thus also capable of
achieving the opposite regime of a second species being immersed
inside a strongly interacting quantum degenerate Fermi gas, similar
to earlier studies in the K-Li mixture \cite{spie09}.

Our experiments with the Yb-Li mixture near a Feshbach resonance demonstrate
effects of an additional species on chemical reaction rates in the
microKelvin regime. We observe a long lifetime for Feshbach molecules, even
in the absence of Pauli blocking. Our demonstrated stability of the
mixture near the unitary regime of the resonance opens various possibilities
of studying strongly interacting fermions immersed in a bath species
or being interrogated by a small probe species. Future experimental
opportunities include realizations of non-equilibrium states, and
studies of superfluid properties, for instance by controlled relative
motion between the two species. Finally, our results constitute an advance in
the manipulation of ultracold mixtures of alkali and alkaline-earth-like atoms \cite{hara11,hans11,baum11}.
These mixtures may be used to produce quantum gases of paramagnetic polar molecules which
are promising for diverse applications such as quantum simulation of lattice spin models \cite{mich06},
tests of fundamental symmetries \cite{huds11}, and probes of time variations in fundamental constants \cite{kaji11}.

We thank Ben Plotkin-Swing for experimental assistance, and J.P. D'Incao,
T.-L. Ho and M.W. Zwierlein for helpful discussions. We gratefully
acknowledge support from the National Science Foundation and the Air
Force Office of Scientific Research. A.K. thanks the NSERC.


\begin{thebibliography}{35}
\bibitem[{odt(2012)}]{odt}
\bibinfo{journal}{The optical trap wavelength $=1064\,$nm, initial power
  $=10\,$W per beam, waist $=26\,\mu$m, crossing angle $=20^{\circ}$. During
  the evaporative cooling step, the power is reduced by a factor of $16$ over
  $3\,$s. At the highest optical trap powers, $\bar{\omega}_{\rm
  Yb}/\bar{\omega}_{\rm Li} = 0.12$ and $U_{\rm Yb}/U_{\rm Li} = 0.45$. At the
  lower final power, gravity reduces the effective trap depth of the heavier
  component more, but has neglible effect on the frequency ratio.}
\bibitem[{bec(2012)}]{bectf}
\bibinfo{journal}{By continuing the forced evaporation to lower trap depths,
  our apparatus can produce pure Yb BECs with $>2\times 10^5$ atoms in
  single-species experiments and deeply Fermi degenerate gases $T_{\rm Li}/T_F
  \simeq 0.1$ in two-species experiments.}
\bibitem[{ybb(2012)}]{ybblast}
\bibinfo{journal}{This step is carried out either immediately before or after
  the field ramp, with no discernible difference in lithium number or
  temperature between the two.}
\bibitem[{processiv(2012)}]{prociv}
\bibinfo{journal}{Since we observe and model process (IV) by its loss signature, we cannot distinguish it from the analogous process resulting in a vibrationally relaxed LiYb molecule, $\mathrm{Li_{2}^{\mathnormal{s}}+Yb}\rightarrow\mathrm{LiYb+Li\hspace{1mm}(+\epsilon_{D})}$. Thus $L'_{2}$ is the combined rate constant for both processes.}
\bibitem[{threebody(2012)}]{threebody}
\bibinfo{journal}{We obtain an upper limit on the rate of processes involving one Li atom and two Yb atoms from lifetime of mixtures of one spin state of Li with Yb at a magnetic field near the Li-Li Feshbach resonance. We obtain an upper limit on the rate of direct three body decay of Li from lifetimes of Li cloud without Yb taken on the BCS side of the resonance. }
\end{thebibliography}
\end{document}